\documentclass[]{aastex701}
\usepackage{hyperref}
 
\begin{document}

\title{Coronal Mass Ejection and Heliospheric Current Sheet Interaction Causing a Long-Duration Magnetic Field Sector Transition }

\author[orcid=0000-0003-4867-7558]{Manuela Temmer}
\affiliation{Institute of Physics, University of Graz, Austria}
\email[show]{manuela.temmer@uni-graz.at}

\author[orcid=0000-0002-2655-2108]{Stephan G. Heinemann}
\affiliation{Institute of Physics, University of Graz, Austria}
\affiliation{Department of Physics, University of Helsinki, Finland}
\email{stephan.heinemann@hmail.at}

\author[orcid=0000-0003-3903-4649]{Nina Dresing}
\affiliation{Department of Physics and Astronomy, University of Turku, Finland}
\email{nina.dresing@utu.fi}

\author[orcid=0000-0002-8680-8267]{Mateja Dumbovic}
\affiliation{Hvar Observatory, Faculty of Geodesy, University of Zagreb, Croatia}
\email{mateja.dumbovic@geof.unizg.hr}

\author[orcid=0000-0002-6998-7224]{Eleanna Asvestari} 
\affiliation{Department of Physics, University of Helsinki, Finland}
\email{eleanna.asvestari@helsinki.fi}

\begin{abstract}
We present a study that combines remote-sensing and in-situ observations of coronal mass ejections (CMEs) interacting with the nearby heliospheric current sheet (HCS). The sequence of eruptive events under study culminates in the largest directly observed flare of solar cycle 25 on 3 October 2024, producing a fast halo CME. Their  source region can be linked to a so-called nested active region (or active longitude) that persisted over several solar rotations. Such long-lived regions reflect deep-seated magnetic structures that shape the global magnetic field configuration.
By applying the drag-based CME propagation model, we connect the near-Sun observations from several CMEs during that activity period with in-situ measurements.
While one of the CMEs propagated on the opposite side of the HCS from Earth, and therefore did not produce in-situ signatures near Earth, we detect, over the period October 5–10, 2024, a complex of HCS and CME structures propagating together with a shock ahead of them. The HCS seems to be locally replaced by the CME signatures, leading to a long-duration sector reversal of more than 48 hours.
This event highlights the intrinsic connection between solar surface structures, the global magnetic field, and the evolution of complex eruptive events.

\end{abstract}


\section{Introduction}

Major energy release in flares---and therefore the occurrence of X-class flares in active regions---is typically linked to the complexity of their magnetic fields \citep[][]{Sammis2000TheSunspots}. Studies have shown that clusters of complex active regions are not uniformly distributed across the Sun but tend to form at preferred longitudes \citep[the so-called active longitudes, active zones, or nested active regions; e.g.,][]{deToma2000AEvolution,Berdyugina2003,Chen2011}. These nests can persist for months to years due to the repeated emergence of magnetic flux and therefore may exert a significant influence on the structure of the solar global magnetic field \citep[][]{Gaizauskas1983,vanDriel2015}.
Recent work reports that the heliospheric current sheet (HCS) can become anchored above nested active regions and, owing to the continual flux emergence, become substantially warped \citep[][]{Finley2024}. Spatial overlaps have also been found between hard X-ray flare locations and Hale sector boundaries \citep[][]{Loumou2018TheLongitudes}, which are segments of the HCS \citep[][]{Svalgaard1976}. More recent results further confirm that active region nests host the most energetic flares \citep[e.g., for solar cycle 24,][]{Finley25}, including those responsible for the strongest geomagnetic storms of solar cycle 25 in May 2024 \citep{Kontogiannis25}.

The large-scale magnetic topology in the solar corona is shaped by the distribution of open and closed magnetic field structures, such as active regions and coronal holes. Coronal mass ejections (CMEs) drive significant changes in this topology. Fast CMEs typically comprise a shock and sheath driven by  magnetic ejecta, which often manifest as flux ropes \citep[see e.g.,][]{Zurbuchen06}. These magnetic structures tend to align with the global magnetic configuration, particularly the orientation of the Heliospheric Current Sheet (HCS) \citep[e.g.][]{Yurchyshyn2001}, or get deflected away from open flux regions such as coronal holes, which are spatially closely related to the HCS structure \citep[][]{cremades06,Gopal2009,Heinemann2019}. The coronal topology and its continuation further out in interplanetary space can get strongly distorted by transient events, namely CMEs \citep[e.g.,][]{Owens2013TheField}. CMEs therefore may have important effects on the HCS itself in terms of shape and local orientation \citep[e.g.,][]{blanco11,Kilpua22,romeo2023}.

In this study, we focus on the magnetic field sector change occurring in in-situ measurements between October 5--9, 2024. Several CMEs released from the Sun between October 1--4, 2024 most likely have interacted with the heliospheric plasma sheet (HPS) surrounding the HCS, leading to a long-duration sector change. We investigate the CME source regions in relation to the HCS to provide a detailed explanation of the structures observed during this period of magnetic sector polarity reversal. We aim to give a consistent picture relating the strongest (as of May 2026) directly observed flare from solar cycle 25 \citep[X9.0 from October 3, 2024; see more details on the flare in][]{Ding25} to the CME and its in-situ measured signals at 1~au. While to some extent it is statistically established that the correlation between CME-related flare intensity and geomagnetic impact is weak \citep[e.g.,][]{Howard2005,Zhang07}, with our analysis we offer an explanation as to why large flares may not produce most strongly geoeffective events.

\section{Data, Methods, and Results} \label{sec:results}

On October 3, 2024 at 12:18 UT, the peak of an energetic X9.0 flare from AR3842 was recorded. It was associated with an eruption that produced a halo CME with the source region location in Stonyhurst heliographic coordinates W03/S20. This event represents the strongest energy release from AR3842, with several CMEs erupting from that region in the days before (including an X7.1 flare) and after, spanning October 1--4, 2024.

The Community Coordinated Modeling Center (CCMC) at NASA provides the publicly available Space Weather Database Of Notifications, Knowledge, Information (DONKI\footnote{\url{ccmc.gsfc.nasa.gov/donki}}). DONKI is meticulously populated and maintained by the Moon to Mars (M2M) Space Weather Analysis office at the Goddard Space Flight Center. There CME 3D parameters (direction, speed, and size) are derived using the Space Weather Prediction Center CME Analysis Tool tool \citep[SWPC-CAT;][]{Millward2013} developed by NOAA and maintained at CCMC. These parameters are typically used as input for CME propagation models. Inspecting additional EUV data from the AIA \citep[Atmospheric Imaging Assembly;][]{Lemen2012} instrument aboard the Solar Dynamics Observatory \citep[SDO;][]{Pesnell2012} together with the DONKI list, we find six CMEs that occurred during October 1–4, 2024\footnote{\url{https://kauai.ccmc.gsfc.nasa.gov/DONKI/search/results?startDate=2024-10-01&endDate=2024-10-04&catalog=M2M_CATALOG&eventType=ALL}}. Table~\ref{tab:CMElist} summarizes the source region, flare association, and appearance time in the LASCO/C2 coronagraph \citep{Brueckner1995} aboard the Solar and Heliospheric Observatory \citep[SOHO;][]{SOHO}. It also includes 3D characteristics of the CMEs, where for certain events, the shock and leading-edge information are provided separately.

\begin{table}[]
 \centering

\begin{tabular}{llll|lrrrr|ll|ll}
       & & &    &               \multicolumn{5}{c}{CME 3D parameters at 21.5~Rs}   & \multicolumn{2}{c}{+DBM} & \multicolumn{2}{c}{Prediction at 1au}    \\
No.     & AR   &  Flare & $t_{\rm LASCO}$  &  $t$  &  $v$   & \textit{lon}  & \textit{lat}  & $ha$  & $v_{\rm SW}$ & $\gamma$ & $t$ & $v$ \\ 
\hline
CME 1$^a$      & 3842 & X7.1 & Oct 1, 23:09              & Oct 2, 02:30      & 960               & $-$10   &$-$12         & 60         & 350      & 0.2     & Oct 4, 12:00     & 520   \\
CME 1$^b$      & \multicolumn{3}{c|}{---}             & Oct 2, 04:20          & 600             & $-$19   &$-$10       & 40         & 350      & 0.2     & Oct 5, 10:35     & 440   \\
CME 2$^b$    & 3842 &  M3.6  & Oct 2, 06:36              & Oct 2, 15:50        & 360              & $-$15   & $-$24        & 25         & 350      & 0.5     & -- & -- \\
CME 3$^b$     & 3842 &  M3.2  & Oct 2, 14:24              & Oct 2, 21:20       & 450              & $-$7  & $-$22         & 20         & 350      & 0.5     & -- & -- \\
CME 4$^a$    & 3842  & X9.0  & Oct 3, 12:48              & Oct 3, 16:30       & 860             & $-$3 &$-$4           & 50         &   350    &  0.2  &   Oct 6, 04:25  &  510 \\
CME 4$^b$    & \multicolumn{3}{c|}{---}                & Oct 3, 16:45         & 820          & 3 &$-$20           & 40         &   350    &  0.2  &   -- & -- \\
\multicolumn{5}{l}{CME 2$^b$+3$^b$+4$^b$ interaction ca.\,at t$_{\rm 105Rs}$ Oct 4, 15:30}     &  500$^*$  & 0     & --     & 50         & 350      & 0.5     & Oct 6, 14:00     & 420   \\
CME 5$^a$    & 3843 & M6.7  & Oct 3, 20:36              & Oct 3, 23:45      & 1080                & 41 & 1          & 50         & 400      & 0.2     & Oct 6, 08:55     & 560   \\
CME 5$^b$    & \multicolumn{3}{c|}{---}                & Oct 4, 02:00        & 650               & 51 & 1         & 50         & 400      & 0.2     &  \textit{Oct 7, 11:15} & \textit{460}   \\
CME 6$^b$   & 3842   &  M4.0    & Oct 4, 05:12              & Oct 4, 11:15   & 560                   & 15 & $-$59         & 30         & 400      & 0.5      & \textit{Oct 7, 17:00}  & \textit{450} \\

\end{tabular}
\caption{M2M catalog CME events from October 1--4, 2024, with shock and bright leading edge treated as separately propagating structures, denoted by indices $^a$ for the shock and $^b$ for the leading edge. Further information covers the CME source region (NOAA active region (AR) number), related flare GOES SXR class, first appearance in LASCO/C2  ($t_{\rm LASCO}$ in UT), and CME 3D parameters from the M2M catalog using SWPC-CAT covering time ($t$ in UT) and speed ($v$ in $\rm km\,s^{-1}$) at a distance of 21.5~Rs, Stonyhurst heliographic longitude (lon in degrees) and latitude (lat in degrees), and half-width (ha in degrees). In addition we extracted from OMNI in-situ data the ambient solar wind speed before CME arrival ($v_{\rm SW}$ in $\rm km\,s^{-1}$) and give the drag parameter $\gamma$ (in $\cdot10^{-7}\mathrm{km^{-1}}$), used as DBM input to calculate the predicted CME arrival time and speed at 1au. $^*$Value for the interaction distance of 105\,Rs.}
    \label{tab:CMElist}
\end{table}

\begin{figure*}[ht!]
\centering
\includegraphics[width=1.\columnwidth]{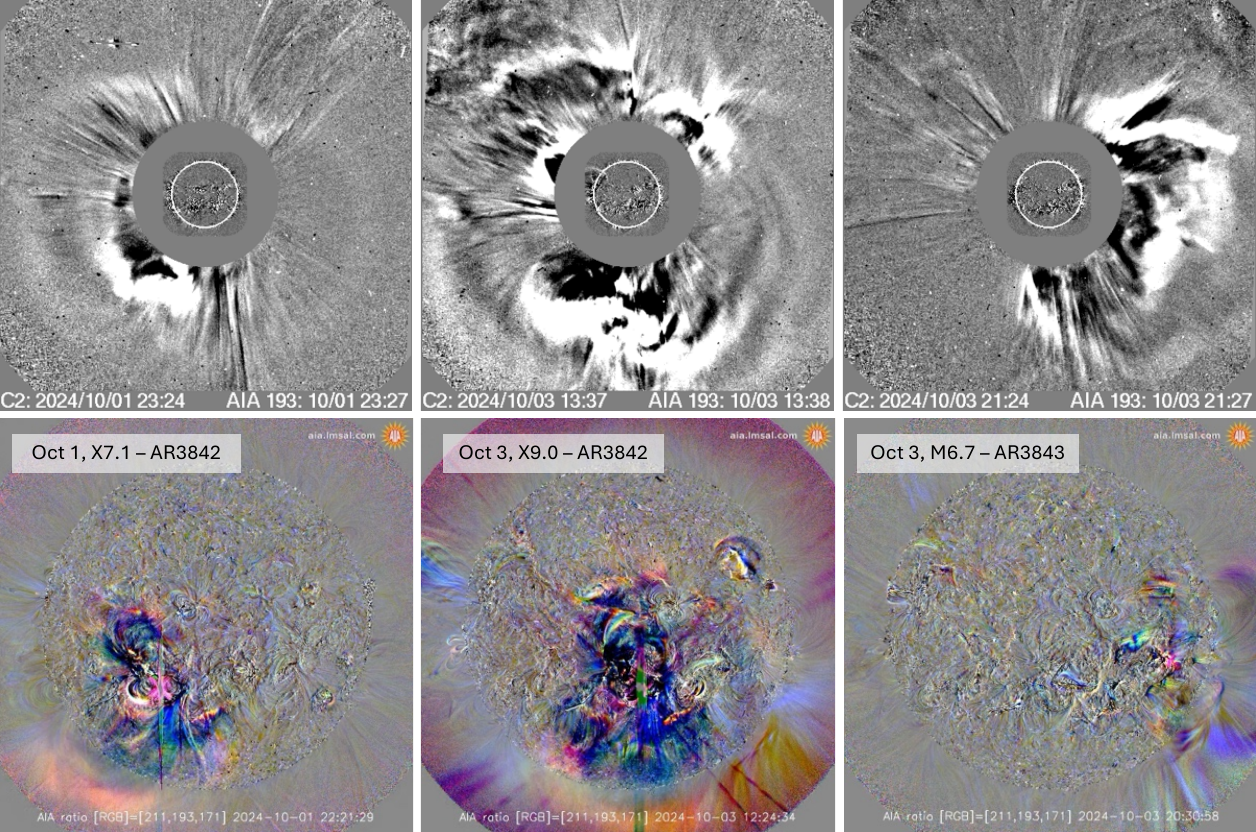}
\caption{Solar surface and white-light structures from the three fastest and most likely Earth-directed CMEs during the activity period October 1--4, 2024. Top: LASCO/C2 white-light and inserted SDO/AIA 193~\AA~EUV running difference images. Bottom: SDO/AIA running ratio EUV images from combined channels of 211~\AA, 193~\AA, and 171~\AA\ filters.} 
\label{fig:flare-CME} 
\end{figure*}

To investigate geoeffective and potentially interacting CMEs, we focus first on the three fastest events listed in Table~\ref{tab:CMElist}: CME1, CME4, and CME5, all of which have shock speeds exceeding 850~$\mathrm{km\,s^{-1}}$. Figure~\ref{fig:flare-CME} shows the solar corona in SDO/AIA EUV three-channel running-ratio images and LASCO/C2 data. CME1 is associated with an X7.1 flare and CME4 is associated with an X9.0 flare. Both are launched from AR3842 and produced coronal waves and large-scale dimming regions, indicative of strongly expanding CMEs. CME5 is associated with an M6.7 flare from AR3843. All CMEs exhibit halo signatures in white-light data. CME1 and CME4, whose source regions are associated with AR3842 near the central meridian, have the highest potential to be detected in situ at Earth.

\begin{figure}[h!]
\centering
\includegraphics[width=0.4\columnwidth]{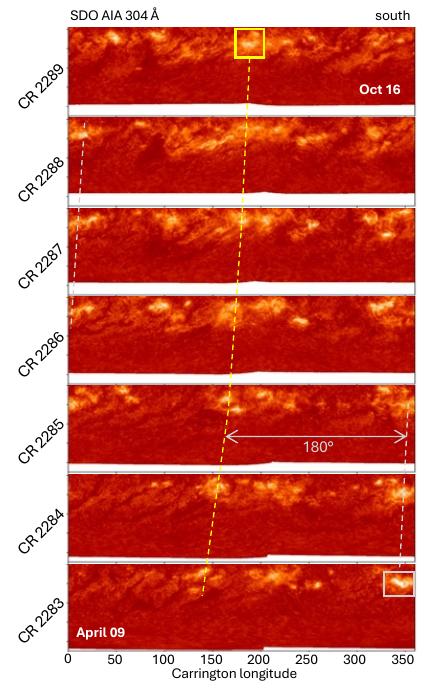}
\caption{Synchronous maps in the 304~\AA~wavelength range for the period  covering CR~2283--2289 (April 9, 2024 - October 16, 2024) using data from the AIA telescope on board SDO. The yellow box marks AR3842 related to the event under study, persisting for several solar rotations (yellow dashed line), hence, can be related to an active longitude or nested active region. The gray box and gray dashed lines refer to the location and further evolution of the nested active region AR3664, being the source of the May 2024 storms. AR3664 and AR3842 are separated in longitude by about $180^{\circ}$.}
\label{fig:ARnest}
\end{figure}

Figure~\ref{fig:ARnest} shows synchronous maps \citep[see][]{Auchere2005} from SDO/AIA 304~\AA\ images of the southern hemisphere, covering the time range April--October 2024. The boxes indicate so-called nested ARs, which are of particular interest. Over that period, we identify two active bands on the Sun that persist over several rotations. One band (yellow dashed line) covers AR3842, which is the main source region of the October 1--4, 2024 events under study (see Table~\ref{tab:CMElist}). The other band (gray dashed line), separated in longitude by about $180^{\circ}$ \citep[see also][]{Usoskin2005PreferredRotation}, is related to the active region producing the so-called May 2024 events, which caused the highly geoeffective Mother's Day or Gannon storm \citep[see e.g., the flash report by][]{Hayakawa2025}. A recent study by \cite{Kontogiannis25}, which examined the magnetic-field evolution of the source region of the May 2024 geomagnetic storms from a $360^{\circ}$ perspective, confirms that it was associated with a nested active region that was supplied by emerging magnetic flux over several months. With that we want to highlight that the source region of the strongest flare observed on the Sun's near-side of solar cycle 25 (as of May 2026) is also linked to a nested active region that persisted for at least 6 months.

\begin{figure*}[ht!]
\centering
\includegraphics[width=.7\columnwidth]{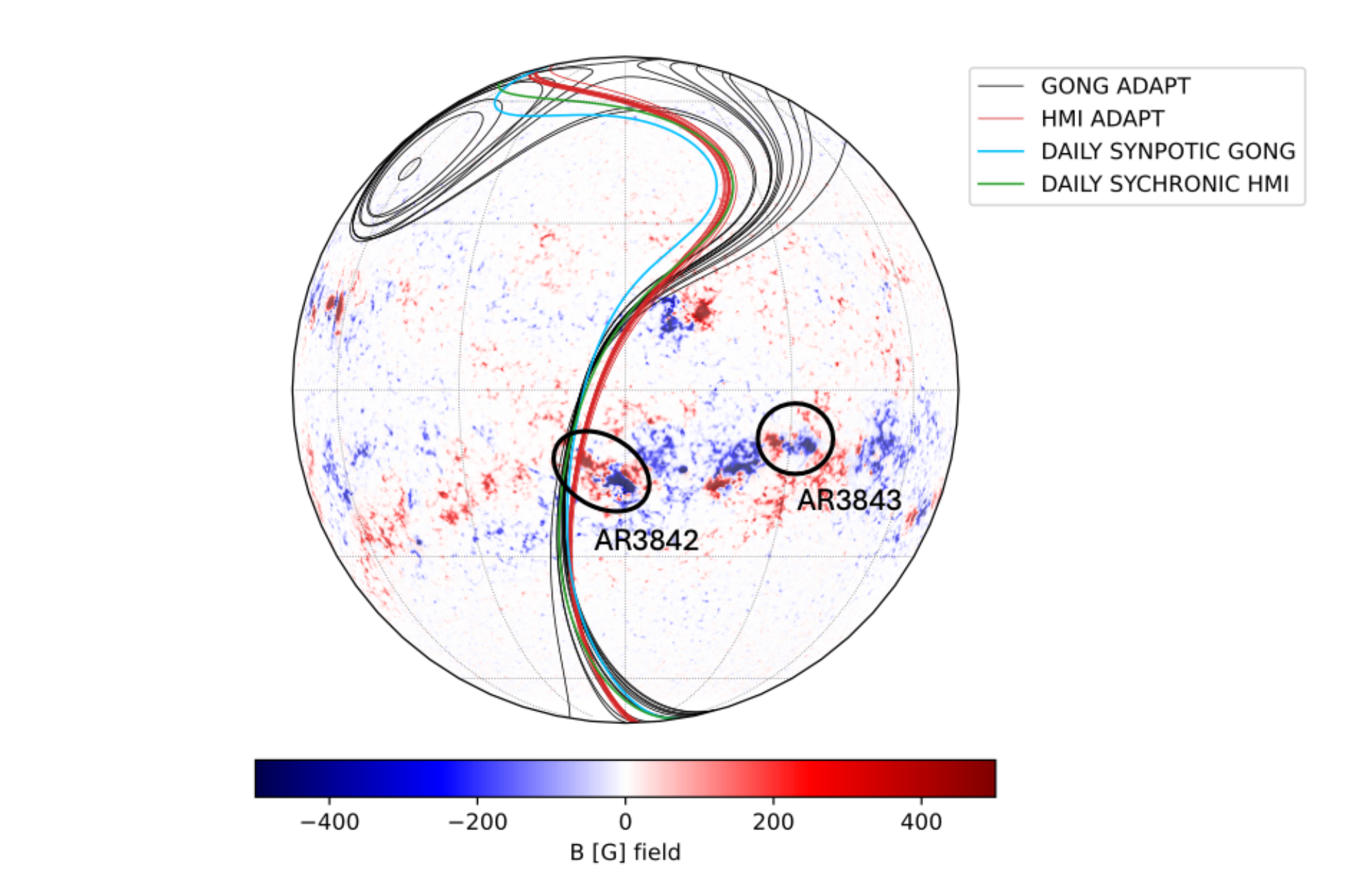}
\caption{PFSS extrapolation using different magnetic field maps (see legend) taken on October 3, 2024, showing the HCS at a source surface height of 2.5~$R_{\odot}$. The active region of interest, AR3842, is lying close to the neutral line of the global magnetic field.}
\label{fig:PFSS} 
\end{figure*}

To infer the relation between that nested active region and the global magnetic field configuration we derive the HCS location. To account for uncertainties, we use different magnetic data sources, namely synoptic HMI \citep[Heliospheric and Magnetic Imager;][]{Schou2012DesignSDO} and GONG \citep[The Global Oscillation Network Group][]{GONG} as well as GONG ADAPT and HMI ADAPT \citep[Air Force Data Assimilative Photospheric Flux Transport ][]{ADAPT} magnetic maps. From these data the potential field source surface was extrapolated using the model \texttt{CIDER}\footnote{\href{https://github.com/jpomoell/cider}{https://github.com/jpomoell/cider}} with a source surface height of 2.5~R$_\odot$. This model uses a staggered finite difference scheme to solve Laplace's equation for the scalar potential, yielding a numerically divergence- and curl-free magnetic field \citep[similar to ][]{Heinemann2026}. Figure~\ref{fig:PFSS} presents the HCS locations as derived from different magnetic maps for October 3, 2024, which show a consistent picture except in the polar regions, where deviations likely result from line-of-sight effects in the magnetic field measurements. We find that AR3842 is closely associated with the HCS. In this context, we investigate how this relationship influences CME evolution and their in-situ characteristics, particularly those originating from AR3842.

\begin{figure}[h!]
\centering
\includegraphics[width=0.8\columnwidth]{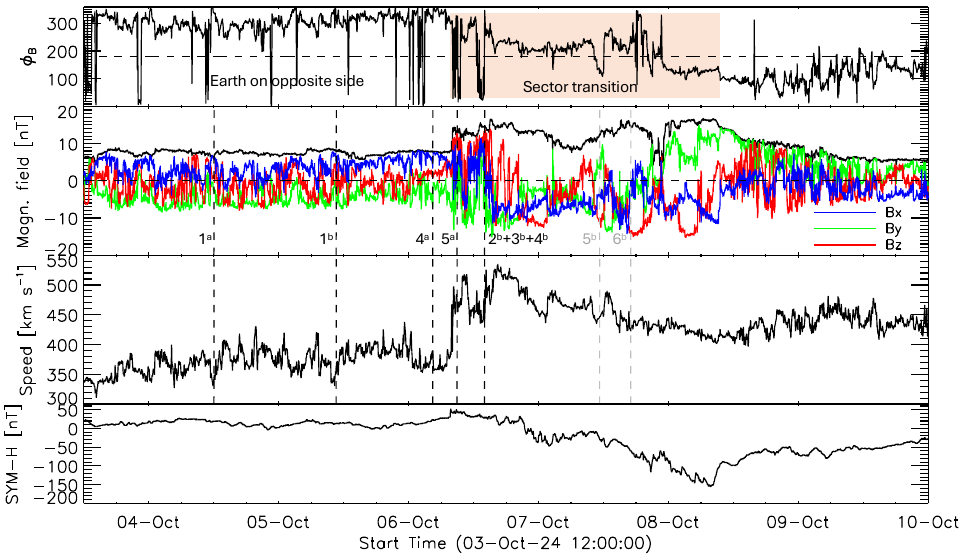}
\caption{OMNI in-situ measurements in GSE coordinates. Top to bottom: azimuthal magnetic field angle ($\phi_B$), total and vector magnetic field, proton bulk speed, and SYM-H index. Vertical dashed lines show DBM results for the predicted arrival times of the CMEs as given in Table~\ref{tab:CMElist}. Predictions for shock ($^a$) and leading edge ($^b$) are given separately. We note that 5$^b$ and 6$^b$ are likely glancing blows for which we marked them in gray. The azimuthal angle shows the change in the magnetic sectors, where we marked the opposite/same location of AR3842 and Earth separated by the HCS.}
\label{fig:context}
\end{figure}

For the in-situ analysis we take measurements from OMNI in GSE (Geocentric Solar Ecliptic) coordinates, as 1-minute data time-shifted to the nose of the Earth's bow shock \citep[][]{king05}\footnote{Data product: OMNI$\_$HRO2$\_$1MIN under \url{https://cdaweb.gsfc.nasa.gov}}. The OMNI dataset includes among others plasma and magnetic field measurements, from which we calculate the magnetic field angles in the azimuthal ($\phi_B$) and polar ($\theta_B$) direction, alpha-to-proton density ratios, as well as magnetic indices, such as the SYM-H index. Figure \ref{fig:context} presents $\phi_B$, the total and vector magnetic field, proton bulk speed, and SYM-H over the time range October 3--10, 2024. Applying the drag-based CME propagation model \citep[DBM;][]{Vrsnak2013,Dumbovic2021_DBM} we infer the arrival times of the structures related to the CMEs. Columns 5--9 in Table~\ref{tab:CMElist} give the CME 3D kinematics and geometries (time at 21.5~Rs, speed at 21.5~Rs, longitude, latitude, and half-width) as taken from the CCMC/M2M catalog. Specific DBM inputs are given in columns 10+11 in Table~\ref{tab:CMElist}, covering the average solar wind speed prior to CME arrival -- consistent with in-situ measurements -- and the drag parameter, $\gamma$, which is chosen to be 0.2$\cdot10^{-7}\mathrm{km^{-1}}$ for CMEs with speeds between 600--1000~$\mathrm{km\,s^{-1}}$ and 0.5$\cdot10^{-7}\mathrm{km^{-1}}$ for CMEs with speeds lower than 600~$\mathrm{km\,s^{-1}}$ \citep[see][]{Vrsnak2014, Calogovic2021}. We note that an informed estimate to be put in DBM is sufficient for our purpose, as we aim to show how things relate between the Sun and in-situ measurements, and not to produce an exact prediction for each CME. Columns 12+13 in Table~\ref{tab:CMElist} give the DBM results covering arrival time and speed. CME1 is expected to be an isolated event, while CME2 and CME3 are interacting at a distance of about 45\,Rs, and then further with CME4 around 105\,Rs. To simulate CME–CME interactions and their subsequent propagation, we adopt the approach described in \cite{Guo2018ModelingMeasurement}, in which the DBM is reinitialized at the CMEs' interaction point using its time and distance as well as average speed with  changing $\gamma$ accordingly. We therefore anticipate the arrival of three closely spaced flux ropes, associated with CME2, CME3, and CME4. This result is consistent with expectations, as all three CMEs originate from the same source region, with each successive eruption exhibiting a higher speed than its predecessor. CME5 and CME6 exhibit longitudes and latitudes indicative of glancing encounters.

We show the derived CME arrival times in Figure~\ref{fig:context} with vertical dashed lines, labeled according to Table~\ref{tab:CMElist}. We separately mark shock and bright leading edge, i.e., related to the magnetic ejecta, because shocks are known to move through the magnetic structure of a CME while flux ropes cannot and form complex regions upon interaction \citep[e.g.,][]{Liu2012InteractionsObservations, Lugaz2015}. Notably we find neither shock nor leading edge of CME1 causing an in-situ signature within $\pm$24 hours of the predicted arrival times October 4--5, 2024 and therefore likely did not reach Earth. The non-detection of CME1 may be explained by the same-side/opposite-side effect \citep[e.g.,][]{Henning1985, Zhao2007, Dumbovic2021, Temmer2025}. This condition is particularly relevant when the HCS is highly inclined, as Earth may cross into a different sector during the CME's propagation time.

A first shock signature is observed on October 6, 2024. We deduce the shock arrival time at Earth from the appearance of the associated geomagnetic sudden storm commencement, which is given at 07:39\,UT. We favor this over the OMNI arrival time (07:50\,UT) because OMNI time-shifting is known to be less reliable for shocks, often resulting in delayed arrival times due to the use of convection-based propagation \citep[see][and the OMNI technical documentation]{king05}. The maximum sheath speed is reached with  $\approx520$~km\,s$^{-1}$. The earliest predicted arrival from the DBM corresponds to the shock of CME4 (4$^a$) on October 6, 2024, at 04:25\,UT, with a speed of 510~km\,s$^{-1}$. This is consistent with observations assuming that the shock is not significantly affected by propagation through preceding CME magnetic structures. Under the same assumption, a likely signature of a shock from CME5 (5$^a$) is predicted to arrive on October 6 at 08:55\,UT with a speed of 560~km\,s$^{-1}$. The interacting magnetic structures from CMEs 2+3+4 (2$^b$+3$^b$+4$^b$) are estimated to arrive on October 6 at 13:50\,UT. The flux ropes associated with CME5 (5$^b$) and CME6 (6$^b$) are expected to produce, if any, only weak signatures, as both are likely glancing encounters based on their propagation directions and geometries, with predicted arrival times between October 7, 11:15 and 13:00\,UT. The delayed in situ signatures of the CMEs, observed up to October 8, 2024, suggest additional deceleration to CME2+3+4, likely caused by interactions with other solar wind structures. The change in the azimuthal angle from $>$180° before the arrival of the disturbance to $<$180° after, indicates that the CMEs are associated with a magnetic sector boundary crossing lasting over 48 hours. In the following, we examine in more detail the relationship between the CMEs with this extended HCS structure.

\begin{figure*}[ht!]
\plotone{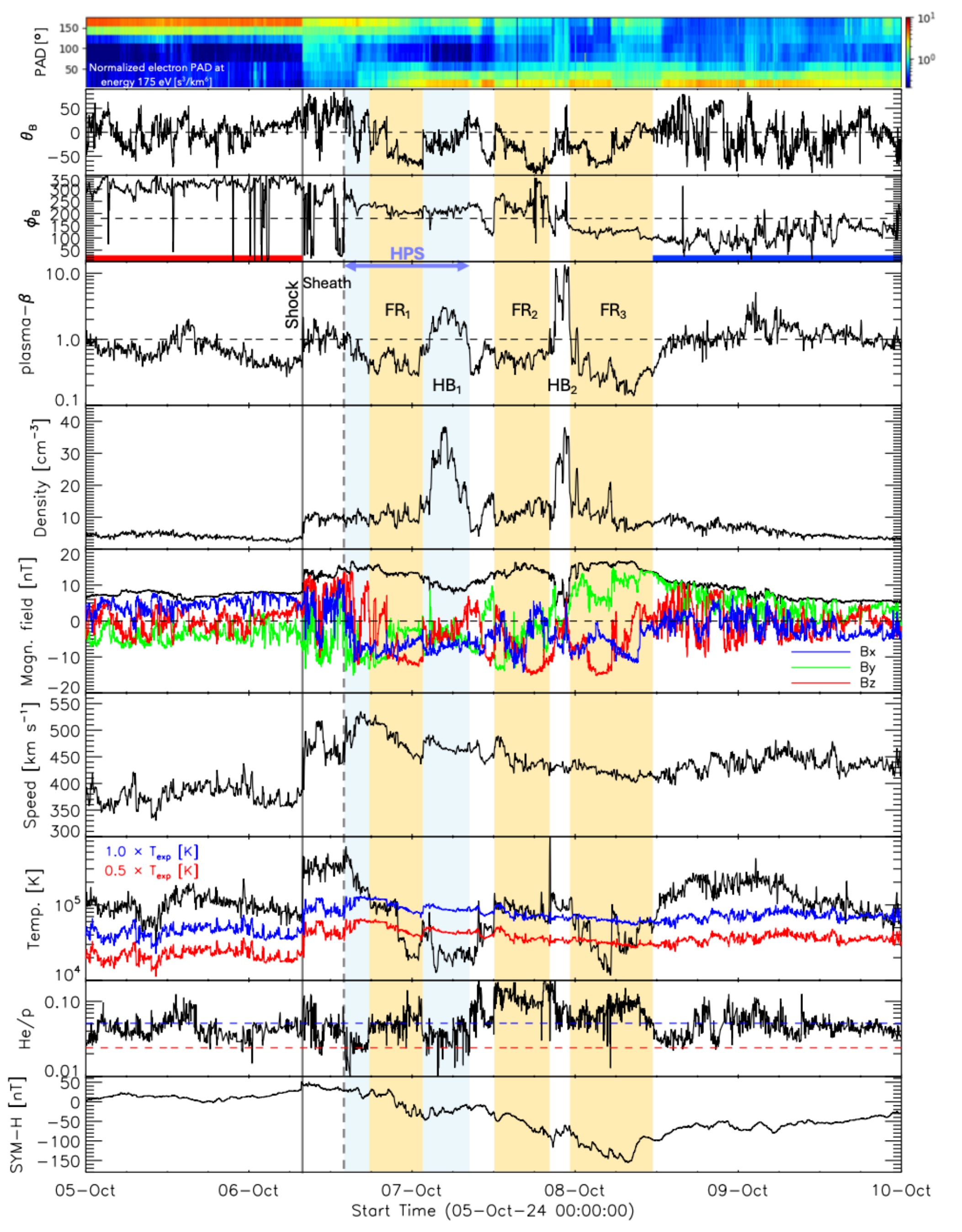}
\caption{In-situ measurements from OMNI in GSE coordinates, covering from top to bottom: electron PADs in the $175$~eV energy range (Wind), polar and azimuthal magnetic field direction ($\theta_B$ and $\phi_B$), plasma-beta, proton density, total magnetic field and vector components, proton bulk speed and temperature, He/p density ratio values, and the SYM-H index. The polarity change in the interplanetary magnetic field according to $B_x$ component, electron PAD and $\phi_B$ is marked in the $\phi_B$ panel with a red-to-blue bar (inward-to-outward magnetic sector). The temperature panel shows the expected temperature ($T_{\rm exp}$) in blue and abnormally low temperature ($0.5\times T_{\rm exp}$) in red, as taken from solar wind statistics \citep[][]{Richardson1995RegionsEjecta}. The He/p panel shows thresholds for magnetic ejecta (0.051) and sector-reversal region plasma (0.024) according to statistics by \cite{Xu15}. FR = flux rope; HPS = heliospheric plasma sheet; HB = high plasma-beta region. } 
\label{fig:insitu}
\end{figure*}

Figure~\ref{fig:insitu} covers OMNI data over the time range October 5--10, 2024. In addition we show electron pitch-angle distributions (PAD) at 175 eV from the 3DP instrument \citep{Lin1995} and 1-minute resolution alpha-to-proton density ratios (He/p) from the L1 spacecraft Wind \citep[][]{Ogilvie1997, Wilson2021}. Low He/p values are found from distinct plasma around the sector reversal region and, according to the statistical values derived by \cite{Xu15}, He/p has a factor of two difference between the average value for ejecta (0.051) and sector reversal plasma (0.024), which we use to distinguish between them. In addition, the proton temperature panel shows the expected temperature ($T_{\rm exp}$) derived from solar wind statistics, where abnormally low solar wind proton temperatures ($<0.5\times$T$_{\rm exp}$) support the identification of flux rope structures \citep[][]{Richardson1995RegionsEjecta}.

We find that the azimuthal angle of the magnetic field is directed approximately along the sunward Parker spiral before the arrival of the shock. Following the shock-sheath, there is a change to $\phi_B\sim$200°, but not a full sector boundary crossing to the outward Parker spiral. As also suggested by the still enhanced temperature of the plasma, the shock-sheath region was not immediately followed by a magnetic flux rope structure. In fact, the low He/p values refer to a sector-reversal region (blue shaded area in Figure~\ref{fig:insitu} starting October 6, $\sim$14~UT), and thus to the associated HCS and heliospheric plasma sheet (HPS) structures. The PAD is isotropic after the shock-sheath part, i.e., over the presumably related HPS structure. It is then followed by a period of a bidirectional electron distribution with a smooth magnetic field and low temperature, as well as enhanced He/p values, all hinting towards a flux rope (FR$_1$; October 6, 17:50~UT -- October 7, 01:40~UT). Subsequently, a high plasma-beta region (HB$_{1}$) is detected with low He/p values and very low proton temperature. In that respect, FR$_1$ could be a small-scale flux rope related to the HPS. Further low plasma-beta, low temperatures, and high He/p values give indication of another FR (FR$_2$; October 7, 12:15~UT--21:10~UT), with a non-uniform rotational profile as derived from the polar angle $\theta_B$, that is followed by another high plasma-beta region (HB$_2$). In comparison to HB$_1$, HB$_2$ reveals high He/p values and higher temperature, hence, the origin of HB$_2$ might be different and not related to the HPS. After HB$_2$ a more pronounced FR (FR$_3$; October 7, 23:00~UT--October 8, 11:30~UT) is identified, overall revealing typical flux rope signatures of high He/p, low plasma-beta, enhanced magnetic field, linearly decreasing speed profile, and low temperature. With a duration of $\approx12.5$ hours FR$_3$ is at the lower limit of the average duration reported for magnetic ejecta \citep[$20.6\pm9.1$~hours;][]{Kilpua2013OnClouds}. The assignment between the CMEs given in Table~\ref{tab:CMElist} and in-situ measurements remains speculative. Assuming that FR$_1$ and HB$_1$ belong to the HPS, a plausible scenario would be that FR$_2$ and FR$_3$ are the compound structures of CME2+3+4, separated by a compression region, HB$_2$. CME2 and CME3 interacted at approximately 45\,Rs, likely forming the complex structure of FR$_2$. This structure was subsequently interacting with the significantly faster CME4 (FR$_3$) at about 105\,Rs. CME5 and CME6 may not be detected, as predicted. When comparing DBM results and start time of FR$_2$, we may infer that due to the interaction between HCS and the magnetic structures of CME2+3+4 (2$^b$+3$^b$+4$^b$; cf.\,Figure~\ref{fig:context}), the CMEs were strongly delayed in their arrival by almost 24 hours. In that scenario the shock from CME4 (and maybe also CME5) is propagating ahead of the entire complex of observed HPS-CME structures. The speed profile shows an overall decline from $530$~km\,s$^{-1}$ to $410$~km\,s$^{-1}$, characterized by a plateau over the high plasma-beta region (HB$_1$) and a monotonic decrease across FR$_2$ and FR$_3$. As indicated by the electron PAD, the magnetic sector reversal begins with the shock arrival on October 6 at 07:39\,UT and evolves over 48 hours. A stable, outward-directed magnetic field polarity is established during flux rope FR$_3$ and persists thereafter.

To evaluate the geomagnetic response, we examine the SYM-H index \citep[][]{SymH}, as shown in the bottom panel of Figure~\ref{fig:insitu}. From that we see a standard storm signature, with a sudden storm commencement at the shock arrival, followed by a continuous decline beginning at the arrival of FR$_1$. The SYM-H index reached its minimum of $-148$nT on October 8, 2024, at 8UT, coinciding with the low-temperature region and the maximum southward magnetic field component within FR$_3$. This can be interpreted as a cumulative effect resulting from the interaction of multiple flux ropes and the extended duration of the disturbance.

\section{Discussion} \label{sec:discussion}

Between October 1--4, 2024, six CMEs as potential Earth-directed events were reported in the CCMC/M2M catalog (see summary in Table~\ref{tab:CMElist}). Among them a CME that is related to the X9.0 solar flare that occurred on October 3, 2024, which is confirmed as the strongest solar flare of Solar Cycle 25 \citep[for details on the solar flare we refer to][]{Ding25}. Along with the intense X7.1 flare-related CME on October 1, 2024, these events originated from the highly active region AR3842 when it was close to the central meridian.

We propagated the six events to 1au using the DBM, modeling the shock and flux rope components separately. Estimated arrival times were found to cover the period October 4--7, 2024. Interestingly, CME1, associated with the X7.1 flare, was predicted as an isolated event but likely did not reach Earth, as no indicative in-situ signatures were detected within $\pm$24 hours of the predicted arrival time October 4--5, 2024. CMEs 2, 3, and 4 (the latter associated with the X9.0 flare) most likely interacted in interplanetary space and propagated toward Earth as a single entity. CMEs 5 and 6 were predicted to be glancing blows and thus likely did not produce significant in-situ signatures. The predicted arrival of CME2+3+4 was temporally associated with the measurement of a crossing of the heliospheric current sheet (HCS). This suggests that the HCS may have isolated Earth from the previously launched CME1, making it an opposite-side event, which is characterized by a lack of both in-situ signals and geomagnetic effects \citep[see also][]{Henning1985, Zhao2007, Dumbovic2021}. While the preceding structures to CME2+3+4 drive modest activity, the strongest southward fields occurring within FR$_2$ and FR$_3$, significantly intensifying the storm finally reaching a SYM-H index of $-$148\,nT on October 8, 2024.

The HCS crossing started simultaneously with the arrival of a shock component October 6, 2024, at 07:39\,UT, as deduced from the appearance of the associated geomagnetic sudden storm commencement. This shock is followed by a period of low $\text{He}/\text{p}$ density ratio, a small-scale flux rope $\text{FR}_1$ (October 6, 17:50\,UT to October 7, 01:40\,UT), and a high plasma-beta region. Subsequently, two additional flux ropes were identified, $\text{FR}_2$ (October 7, 12:15--21:10\,UT) and $\text{FR}_3$ (October 7, 23:00\,UT to October 8, 11:30\,UT), separated by another high plasma-beta region. These results are generally consistent with existing catalogs. The manually maintained list by \cite{RClist}, based on OMNI data, reports the same shock arrival time for the sudden storm commencement on October 6 at 07:39\,UT, followed by two consecutive CMEs: the first magnetic ejecta spanning October 6, 14:00\,UT to October 7, 20:00\,UT, and a second one from October 8, 00:00--11:00\,UT. The Helio4Cast catalog \citep{Helio4cast_Moestl}, which employs automated detection of flux rope signatures (enhanced magnetic field strength and rotation), reports a shock arrival in the Wind data on October 6 at 07:08\,UT, followed by three magnetic ejecta: (1) October 6, 15:02\,UT to October 7, 00:56\,UT, (2) October 7, 11:36--20:20\,UT, and (3) October 7, 22:10\,UT to October 8, 08:44\,UT.

\begin{figure}
    \centering
    \includegraphics[width=0.8\linewidth]{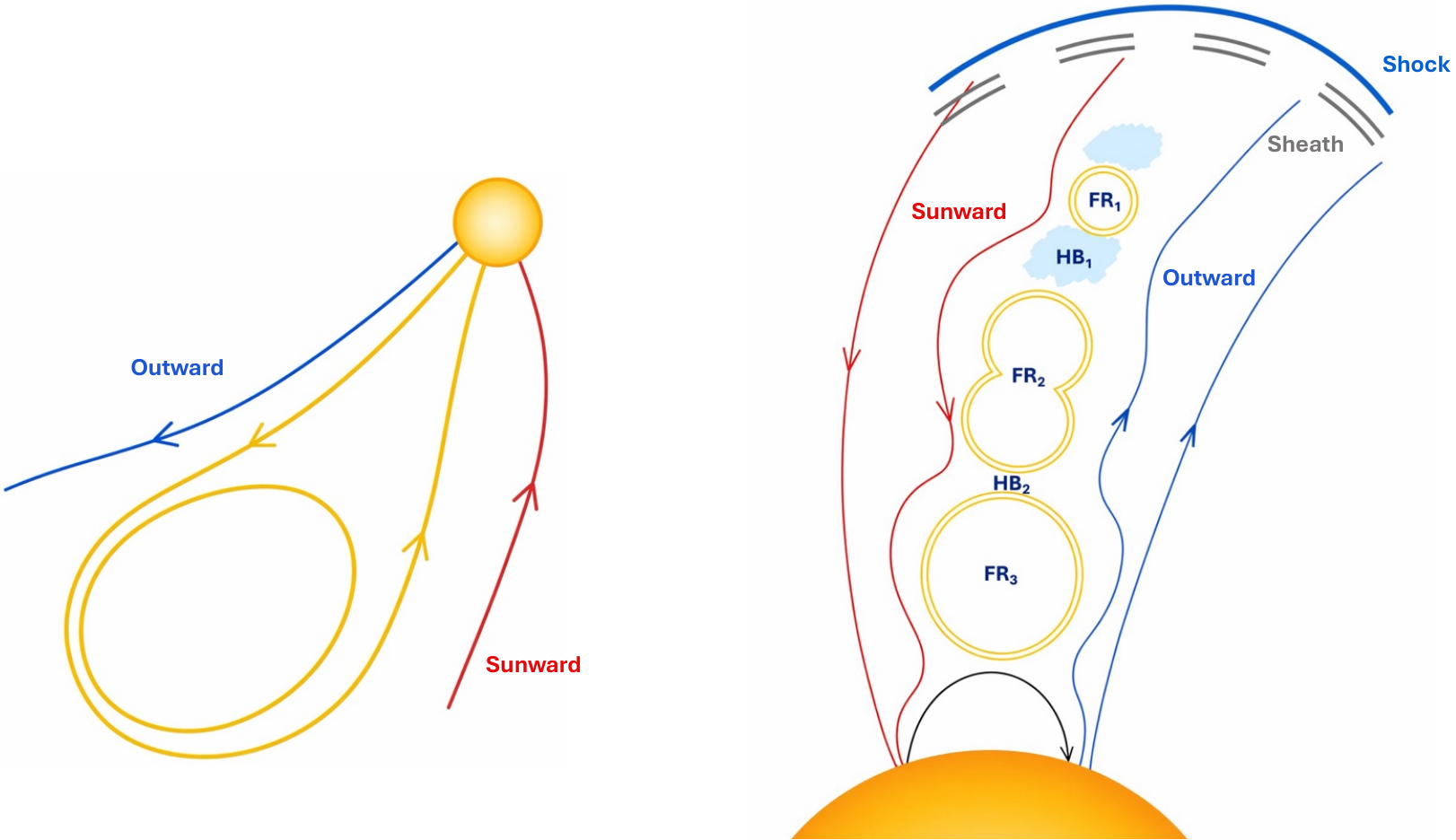}
    \caption{Left: Replacement or widening of the HCS by an expanding closed magnetic structure within the HCS \citep[adapted from][]{Crooker2006}. Right: Illustration of sequence of shock, sheath, flux ropes (FR), and high plasma-beta structures (HB), in between a changing magnetic sector, with blue-shaded areas marking HPS-related structures (cf.\,Figure~\ref{fig:insitu}).}
    \label{fig:cartoon}
\end{figure}

However, our study suggests that $\text{FR}_1$ is a heliospheric plasma sheet (HPS)-related structure surrounding the HCS \citep[e.g.,][]{Lavraud2020}. Besides the in-situ measurements, this is supported by the close association between the source region of the CMEs and the HCS, as indicated by the PFSS extrapolation. We may speculate that FR$_2$, showing a non-uniform rotation in the polar magnetic field angle, might be related to the interacting CMEs 2 and 3. Presumably, FR$_2$ interacting  further out with the faster CME4, i.e., FR$_3$, the flux ropes are separated by a strong compression region in between (HB$_2$). The CME-HCS/HPS complex as such becomes a long-duration magnetic field sector transition event covering more than 48 hours between October 6--8, 2024. Previous studies have discussed cases where the HCS does not drape around a CME but rather the CME locally replaces the HCS, resulting in a long-lasting sector reversal \citep[see][]{Forsyth1997, Crooker2006}. Figure~\ref{fig:cartoon} illustrates the scenario of the successive CMEs erupting under the HCS structure (left), and replacing the HCS by a closed flux rope structure (right).

Signatures of the HCS detected within CME structures have been investigated primarily in the CME sheath region, where occasional reconnection signatures point to ongoing interaction processes \citep[][]{Ala-Lahti25}. Due to the draping of the upstream interplanetary magnetic field over the leading edge of the magnetic ejecta, strong compression occurs \citep[see also][]{Jones02}, which may result in the ingestion of the warped HCS into the CME sheath \citep[][]{Kilpua22}. Interaction between the HCS and CME magnetic ejecta has also been observed by multiple spacecraft at different heliocentric distances, revealing changes in the magnetic-field structure induced by the HCS encounter \citep[][]{Winslow2016}. Recent work by \cite{romeo2023} and \cite{Liu24} demonstrates that interactions between CMEs and the HCS result in global restructuring and increased warping of the current sheet.

In this context, we want to emphasize our holistic approach for this study. By examining the source region of the CME, AR3842, our analysis reveals that this region forms part of a nested active region, or active longitude, which has persisted for at least six months. This nest appears to be separated in longitude by roughly $180^{\circ}$ from a second active longitude responsible for the highly geoeffective May 2024 events \citep[see][]{Kontogiannis25}. Also during solar cycle 23, two active longitudes hosting superactive regions in both hemispheres were reported, separated by about $160^{\circ}$ to $200^{\circ}$ \citep[e.g.,][]{Chen2011}. The same authors found that these ``superactive'' regions generate more than $40\%$ of all X-class flares in a solar cycle and that all such regions have the potential to produce fast CMEs capable of driving strong geomagnetic storms. While active longitudes tend to produce high-energy flares and are often associated with strong, magnetically driven CMEs \citep[flare–CME feedback;][]{Zhang2006,Temmer2008}, they can also significantly shape the global coronal magnetic field \citep[see][]{Finley2024}. Indeed, from magnetic field extrapolations we find that AR3842 hosts the global neutral line, i.e., the heliospheric current sheet (cf.\,Figure~\ref{fig:PFSS}). Consequently, CME-HCS interaction--or at least, their evolution into complex interplanetary magnetic structures--becomes essentially unavoidable during maximum activity phase.

\section{Conclusion}

The solar activity period October 1--4, 2024 caused enhanced geomagnetic activity over October 5--10, 2024 and was driven by a complex ``CME-HCS/HPS compound structure'' related to the October 3, 2024 event. This complex covers a sequence of flux ropes and high plasma-beta structures, which eventually replaced or widened the HCS, creating a rare, long-duration magnetic sector transition event lasting over 48 hours. Notably, the CME launched October 1, 2024 did not reach Earth as it propagated on the opposite side of the HCS. The study offers a novel holistic perspective by linking such complex events to long-lived nested active regions that persist for months and produce recurrent, highly geoeffective solar events. The study underscores the need to further investigate the dynamic evolution of HCS and HPS-related structures from their active source regions into the heliosphere.

As solar active longitudes and their behaviour are found to be similar to stellar activity phenomena \citep[][]{Berdyugina1998}, these results may have broader implications for better understanding the flare-CME relation on magnetically active stars \citep[e.g.,][]{Koller2021SearchSpectra}.

\begin{acknowledgments}
We thank the anonymous referee for the valuable comments, which helped improve the study. MT and MD acknowledge the support from the Austrian-Croatian Bilateral Scientific Project ``Analysis of solar eruptive phenomena from cradle to grave''. SGH acknowledges funding from the Austrian Science Fund (FWF) Erwin-Schr\"odinger fellowship [10.55776/J4560] and funding from the Research Council of Finland (Academy Fellowship) [370747; RIB-Wind]. ND is grateful for support by the Research Council of Finland (SHOCKSEE, grant No.\ 346902, and AIPAD, grant No.\ 368509). EA acknowledges support from the Research Council of Finland (Research Fellow grant number 355659). MD acknowledges support from the European Union – NextGenerationEU within the framework of the National Recovery and Resilience Plan (NPOO), project “Eruptive processes on the Sun”. We acknowledge funding from the European Union's Horizon Europe research and innovation programme under grant agreement No.\ 101134999 (SOLER). The paper reflects only the authors' view and the European Commission is not responsible for any use that may be made of the information it contains. The authors acknowledge the financial support by the University of Graz.
\end{acknowledgments}

\begin{contribution}
MT initiated the study, performed the main part of the analysis and wrote the manuscript. SH provided the analysis for the magnetic field extrapolation. ND and EA provided expertise and tools to prepare and interpret 3DP data. All authors discussed and proofread the manuscript.


\end{contribution}

%



\bibliographystyle{aasjournalv7}

\end{document}